\documentclass{article}
\usepackage[utf8]{inputenc}
\usepackage{caption}
\usepackage{dsfont}
\usepackage{amsmath}
\usepackage[caption=false,font=footnotesize]{subfig}
\usepackage{xcolor}
\usepackage[margin=1.25in]{geometry}

\usepackage{enumitem,amssymb}
\usepackage{pifont}

\usepackage{siunitx}
\usepackage[numbers]{natbib}
\usepackage{graphicx}

\title{ A rewriting of the relation between the acolinearity of annihilation photons and their energy in the context of positron emission tomography }
\author{Maxime Toussaint$^{1}$, Francis Loignon-Houle$^{2}$, Étienne Auger$^{3}$, \and Jean-Pierre Dussault$^{4}$ and Roger Lecomte$^{1}$}
\date{%
    $^1$ Sherbrooke Molecular Imaging Center of CRCHUS and Department of Nuclear Medicine and Radiobiology, Université de Sherbrooke, Sherbrooke, QC, Canada\\%
    $^2$ Instituto de Instrumentación para Imagen Molecular (I3M), Centro Mixto CSIC - Universitat Politècnica de València, Valencia, Spain\\
    $^3$ IR\&T, Sherbrooke, QC, Canada\\
    $^4$ Department of Computer Science, Université de Sherbrooke, Sherbrooke, QC, Canada\\[2ex]%
    \today
}

\begin{document}

\maketitle

    \section{ Context }
\begin{figure}
            \centering
            \includegraphics[width=0.45\textwidth]{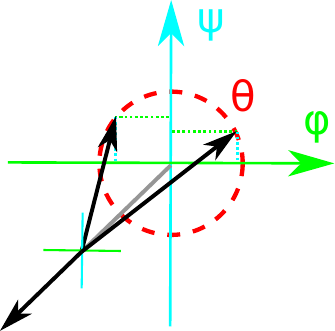}
            \caption{%
                Representation of two examples of acolinearity that can occur in a PET annihilation, described with their amplitude ($\theta$, in red) and their deviation ($(\phi, \psi)$, in green and cyan).
                Note that the two examples have the same amplitude but different deviation. 
                The gray line represents the colinear case.
            }
            \label{fig:amplitudeVsDirection}
        \end{figure}

        In the literature, the acolinearity of the annihilation photons observed in Positron Emission Tomography (PET) is described as following a Gaussian distribution~\cite{levin1999calculation,espana2009penelopet}. 
        However, it is never explicitly said if it refers to the amplitude of the acolinearity angle or its 2D distribution relative to the case without acolinearity (herein defined as the acolinearity deviation) (see Figure~\ref{fig:amplitudeVsDirection}).
        Since the former is obtained by integrating the latter, a wrong interpretation would lead to very different results.
        The blur induced by acolinearity in the image space is assumed to be a 2D Gaussian distribution~\cite{levin1999calculation}, which corresponds to the acolinearity deviation following a 2D Gaussian distribution.

        The study of acolinearity was carried out for various materials using collimation and distance-based setups~\cite{lang1957angular,debenedetti1950angular,colombino1960angular,colombino1963point,colombino1964angular}.  
        For PET, it was shown that acolinearity followed a Gaussian distribution~\cite{colombino1965study}.
        However, the terms used in that article leave enough leeway that both interpretations, amplitude or deviation, would work. 
        To the best of our understanding, the experiment samples the acolinearity distribution over a 1D line which would mean that acolinearity follows a 2D Gaussian distribution. 

        The paper of Shibuya et al., see~\cite{shibuya2007annihilation}, differs from the previous studies since it is based on the precise measurement of the energy of the annihilation photons.
        They also show that acolinearity follows a Gaussian distribution in the context of PET.
        However, their notation, which relies on being on the plane where the two annihilation photons travel, could mean that their observation refers to the amplitude of the acolinearity angle.
        If that understanding is correct, it would mean that acolinearity deviation follows a 2D Gaussian distribution divided by the norm of its argument.  
        This would contradict the conclusion made in the studies mentioned previously.
        Thus, we decided to revisit the proof presented in~\cite{shibuya2007annihilation} by using an explicit description of the acolinearity in the 3D unit sphere.

        The following section describes these details.
        It confirms that acolinearity deviation follows a 2D Gaussian distribution.

    \section{ 3D version of the proof made in Shibuya et al. (2007) }
    \label{sec:preuveConcluShibuya3D}    
        In this section, we replicate the proof made in~\cite{shibuya2007annihilation} except that the momentum is represented as a 3D vector.
        In short, we add the $z$ axis, and thus defining the momentum in 3D and adapt the proof of~\cite{shibuya2007annihilation} accordingly.
        While we add more details in the equations development just to make sure we are not missing anything, we did not copy the explanation of the approximations and equations of~\cite{shibuya2007annihilation}.
        Also, references to equations from~\cite{shibuya2007annihilation} will be colored in green to make a distinction with the equations developed here.

\begin{figure}
            \centering
            \includegraphics[width=0.45\textwidth]{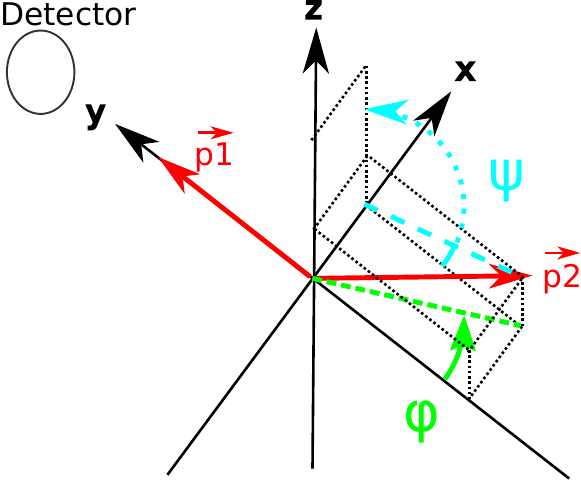}
            \caption{%
                The 3D coordinate system used to parameterize the experiment described in~\cite{shibuya2007annihilation}.
            }
            \label{fig:systAxeShibya3D}
        \end{figure}

        Now, let us define the notations, which are visually represented in Figure~\ref{fig:systAxeShibya3D}.
        Let $\vec{p_1}$ and $\vec{p_2}$ be the momentum of the two annihilation photons where $\vec{p_1}$ corresponds to the annihilation photon that was detected.
        Note that $p_1$ and $p_2$ will be used to respectively define the norm of $\vec{p_1}$ and $\vec{p_2}$.
        Here, the parameterization of $\vec{p_2}$ uses two angles.
        $\phi$ defines the counter-clockwise rotation in the plane $xy$ with $\phi = 0$ being the opposite of the $y$ axis.
        $\psi$ represents the clockwise rotation in the plane $yz$ with $\psi = 0$ being the inverse of the $y$ axis.
        We assume that the acolinearity amplitude ($\theta$) is small enough that $\sin(\theta) \approx \theta$ and, as such, $\phi$ and $\psi$ can be treated as cartesian coordinates.
        In order to make the coordinate system deterministic for a given $\vec{p_1}$, let the $x$ axis be perpendicular to the gravity vector and the $z$ axis have an obtuse angle with the gravity vector.
        
        Following the system defined in Figure~\ref{fig:systAxeShibya3D}, we have:
        $$ \vec{p_1} = [0, p_1, 0] $$
        and
        $$ \vec{p_2} = [p_2 \sin(\phi) \cos(\psi), - p_2 \cos(\phi) \cos(\psi), p_2 \sin(\psi)]. $$
        By the law of momentum conservation, which is $\vec{p_1} + \vec{p_2} = \vec{p}_{e-p}$, we have that
        \begin{equation}
            \label{eq:conserMom_3D}
            \vec{p}_{e-p} = [p_x, p_y, p_z] = [p_2 \sin(\phi) \cos(\psi), p_1 - p_2 \cos(\phi) \cos(\psi), p_2 \sin(\psi)].
        \end{equation}

        The law of conservation of energy, see equation {\color{green}(3)} in~\cite{shibuya2007annihilation}, implies:
        \begin{equation}
        \label{eq:conserEner_3D}
            2 m_0 c^2 = c p_1 + c p_2
        \end{equation}
        where $m_0$ is the mass of one electron and $c$ the speed of light in the vacuum.

        As described in~\cite{shibuya2007annihilation}, the number of equations is equal to the number of variables ($p_1$, $p_2$, $\theta$) which means that if any of the three variables is known, we can compute the other two.
        The authors of~\cite{shibuya2007annihilation} measure $c p_1$ with their setup.
        
        First, we isolate $p_2$
        \begin{align}
            p_y  &= p_1 - p_2 \cos(\phi) \cos(\psi) & \text{From } \eqref{eq:conserMom_3D} \nonumber \\
            p_2  &= \frac{p_1 - p_y}{\cos(\phi) \cos(\psi)} & \label{eq:shibuyaRedo_5upper_intermediaire_3D}
        \end{align}             

        We define $p_1$ relative to $p_y$ which is the first part of equation {\color{green}(5)} in~\cite{shibuya2007annihilation}.
        \begin{align}
           c p_1  &= 2 m_0 c^2 - c \left(\frac{p_1 - p_y}{\cos(\phi) \cos(\psi)}\right) & \text{Substitution of } \eqref{eq:shibuyaRedo_5upper_intermediaire_3D} \text{ in }  \eqref{eq:conserEner_3D} \nonumber \\
            c p_1 + \frac{c p_1}{\cos(\phi) \cos(\psi)}  &= 2 m_0 c^2 + c \left(\frac{p_y}{\cos(\phi) \cos(\psi)}\right) &  \nonumber \\
            c p_1 \left(\frac{\cos(\phi) \cos(\psi) + 1}{\cos(\phi) \cos(\psi)}\right) &= 2 m_0 c^2 + c \left(\frac{p_y}{\cos(\phi) \cos(\psi)}\right) & \nonumber \\
            c p_1 &= \frac{2 m_0 c^2 \cos(\phi) \cos(\psi) + c p_y}{1 + \cos(\phi) \cos(\psi)} &  \nonumber \\
                    &\approx m_0 c^2 + 0.5 c p_y & \text{With } \cos(\phi) \approx 1.0, \cos(\psi) \approx 1.0 \label{eq:shibuyaRedo_5upper_3D}
        \end{align}  
        
        From~\eqref{eq:conserMom_3D}, we have 
        \begin{align}
            p_1 &= p_y + p_2 \cos(\phi) \cos(\psi) & \label{eq:shibuyaRedo_5lower_intermediaire_3D}
        \end{align}             

        We define $p_2$ relative to $p_y$ which is the second part of equation {\color{green}(5)} in~\cite{shibuya2007annihilation}.
          \begin{align}
           c p_2  &= 2 m_0 c^2 - c \left(p_y + p_2 \cos(\phi) \cos(\psi) \right) & \text{Substitution of } \eqref{eq:shibuyaRedo_5lower_intermediaire_3D} \text{ in }  \eqref{eq:conserEner_3D} \nonumber \\
            c p_2 + c p_2 \cos(\phi) \cos(\psi)  &= 2 m_0 c^2 - c p_y &  \nonumber \\
            c p_2 &= \frac{2 m_0 c^2 - c p_y}{1 + \cos(\phi) \cos(\psi)} &   \nonumber \\
                    &\approx m_0 c^2 - 0.5 c p_y & \text{With } \cos(\phi) \approx 1.0, \cos(\psi) \approx 1.0 \label{eq:shibuyaRedo_5lower_3D}
        \end{align} 

        In~\cite{shibuya2007annihilation}, the authors would then define $p_x$ relative to $\theta$, resulting in equations {\color{green}(6)} and {\color{green}(7)}.
        Here, we define $p_{\perp} := \sqrt{p_x^2 + p_z^2}$ relative to $\sqrt{\phi^2 + \psi^2}$.
         \begin{align}
           p_{\perp}  &= \sqrt{p_2^2 (\sin^2(\phi) \cos^2(\psi) + \sin^2(\psi))} & \text{From } \eqref{eq:conserMom_3D} \nonumber \\
                &= p_2 \sqrt{\sin^2(\phi) + \sin^2(\psi)} & p_2 > 0.0, \cos^2(\psi) \approx 1.0 \nonumber \\
                &\approx (m_0 c - 0.5 p_y) \sqrt{\sin^2(\phi) + \sin^2(\psi)} & \text{From } \eqref{eq:shibuyaRedo_5lower_3D} \nonumber \\
                &\approx (m_0 c - 0.5 p_y) \sqrt{\phi^2 + \psi^2} & \text{With } \sin(\phi) \approx \phi, \sin(\psi) \approx \psi  \label{eq:shibuyaRedo_6_3D} \\
                &\approx m_0 c \sqrt{\phi^2 + \psi^2} & \text{With } |p_{\perp}| \gg 0.5 \sqrt{\phi^2 + \psi^2} |p_y|  \label{eq:shibuyaRedo_7_3D_pPerp}
        \end{align}    
        
        From here, the authors of~\cite{shibuya2007annihilation} speak in terms of distribution. 
        Let $\langle X \rangle$ be the distribution of the variable $X$.
        We define $\Delta E$ relative to $p_y$ where $\Delta E = m_0c^2 - cp_1 = -(m_0c^2 - cp_2)$.
        From~\eqref{eq:shibuyaRedo_5upper_3D} and~\eqref{eq:shibuyaRedo_5lower_3D}, we thus have
        \begin{equation}
        \label{eq:shibuya_DeltaE_3D}
            \langle \Delta E \rangle \equiv 0.5 c \langle p_y \rangle,
        \end{equation}   
        which is equation {\color{green}(8)} from~\cite{shibuya2007annihilation}.
        For equation {\color{green}(9)} of \cite{shibuya2007annihilation} (Note: the position of the term $m_0 c$ is wrong in the paper), we have, by using equation~\eqref{eq:shibuyaRedo_7_3D_pPerp}, the following
        \begin{equation}
            \label{eq:shibuya_theta_3D_pPerp}
            m_0 c \left\langle \sqrt{\phi^2 + \psi^2} \right\rangle \equiv \left \langle \sqrt{p_x^2 + p_z^2} \right \rangle.
        \end{equation}   
        
        In~\cite{shibuya2007annihilation}, the authors claim that $p_x$ and $p_y$ can be considered statistically equivalent in the human body since the molecules are not oriented in a certain direction. 
        Thus, $\langle p_x \rangle \equiv \langle p_y \rangle \equiv \langle p_z \rangle$ and we have 
        \begin{equation}
            \label{eq:shibuya_conclu_3D_pPerp}
            \left\langle \sqrt{\phi^2 + \psi^2} \right\rangle \equiv \frac{2 \sqrt{\langle \Delta E \rangle^2 + \langle\Delta E \rangle^2} }{m_0 c^2}.
        \end{equation}   
        Following the previous assumption, we can infer that $\langle \phi \rangle \equiv \langle \psi \rangle$. 
        If the location parameter of $\langle \Delta E \rangle$ is zero, we can conclude that $\langle \phi \rangle \equiv \langle \Delta E \rangle$ et $\langle \psi \rangle \equiv \langle \Delta E \rangle$.
        Since $\langle \Delta E \rangle$ is a Gaussian distribution centered in 0, we can thus conclude that $\langle \phi \rangle$ and $\langle \psi \rangle$ are both Gaussian distribution centered at 0. 
        Thus, the distribution of acolinearity deviation in PET is a 2D Gaussian distribution centered at (0, 0) in the coordinate system presented in Figure~\ref{fig:systAxeShibya3D}.

\bibliographystyle{unsrt}
\bibliography{main}

\end{document}